\documentstyle[prl,aps,floats,psfig]{revtex}

\begin{document}
\widetext
\draft
\twocolumn[\hsize\textwidth\columnwidth\hsize\csname @twocolumnfalse\endcsname
\title{Library Design in Combinatorial Chemistry by Monte Carlo Methods}

\author{Marco Falcioni and Michael W.\ Deem}

\address{Chemical Engineering Department,
University of California,
Los Angeles, CA, 90095--1592
}

\maketitle

\begin{abstract}
Strategies for searching the space of variables in 
combinatorial chemistry experiments are presented, and
a random energy model of combinatorial chemistry experiments
is introduced.  The search strategies, derived by 
analogy with the computer modeling technique of Monte Carlo,
effectively search the variable space even in
combinatorial chemistry experiments of modest size.
Efficient implementations of the library design and redesign
strategies are feasible with current experimental capabilities.
\end{abstract}


]

\newpage

\section{Introduction}
 The goal of combinatorial materials discovery
 is to find
compositions of matter that maximize a specific material
property \cite{2a,15,3}, such as superconductivity \cite{4},
magnetoresistance \cite{22}, luminescence \cite{11,5,10}, 
ligand specificity \cite{23},
sensor response \cite{24}, or catalytic
activity \cite{15,16,31,18,17,21,7}. This problem can be reformulated
as one of searching a multi-dimensional space, with the material
composition, impurity levels, and synthesis conditions as variables.
The property to be optimized, the figure of merit, is generally an
unknown function of the variables and can be measured only
experimentally.  

Present approaches to combinatorial library design and screening
invariably perform a grid search in composition space, followed by a
``steepest-ascent'' maximization of the figure of merit.  This
procedure becomes inefficient in high-dimensional spaces or when
the figure of merit is not a smooth function of the variables, and its use
has limited most combinatorial
chemistry experiments to ternary or quaternary compounds.

In this paper, we suggest new experimental
protocols for searching the space of variables in combinatorial
chemistry, exploiting an analogy between combinatorial materials
discovery and Monte Carlo computer modeling methods.
In Section II we discuss several of these strategies for library design
and redesign.
In Section III we introduce the Random Phase Volume Model
that we will use to compare the different methods.
The effectiveness of different strategies is discussed in
Section IV. We conclude in Section V.

\section{Sampling the Space of Variables in Materials Discovery}

Several variables can be manipulated in order to seek the
material with the optimal figure of merit.  Material composition is
certainly a variable.  But also, film thickness \cite{8} and deposition
method \cite{Novet95}
are variables for materials made in thin film form. The
processing history, such as temperature, pressure, pH, and atmospheric
composition, is a variable.  The guest composition or impurity level
can greatly affect the figure of merit \cite{7}. In addition, the
``crystallinity'' of the material can affect the observed figure of
merit \cite{8}. Finally, the method of nucleation or synthesis may
affect the phase or morphology of the material and so affect the
figure of merit \cite{Zones98}.

We assume that  the composition and
non-composition variables  of each
sample can be changed independently \cite{2a,17}.
Then, instead of a grid search on the
composition and non-composition variables,
we consider choosing the
variables \emph{at random} from the allowed values.  We also consider
choosing the variables in a fashion that attempts to maximize the amount
of information gained from the limited number of samples screened,
{\em via} a quasi-random, low-discrepancy sequence \cite{LDS1}. 

We further consider performing multiple rounds of screening, 
incorporating feedback as the experiment proceeds by
treating the
combinatorial chemistry experiment as a Monte Carlo in the laboratory.
This leads to sampling the experimental figure of merit, $E$,
proportional to $\exp(\beta E)$.
If $\beta$ is large, then the Monte Carlo procedure
will seek out values of the composition and non-composition variables
that maximize the figure of merit.  If $\beta$ is too large, however,
the Monte Carlo procedure will get stuck in relatively low-lying local
maxima.  The first round is initiated by choosing the 
composition and non-composition
variables at random from the allowed values.
The variables are changed in succeeding rounds as dictated by the Monte Carlo
procedure.

\begin{figure}[htbp]
\centering
\leavevmode
\psfig{file=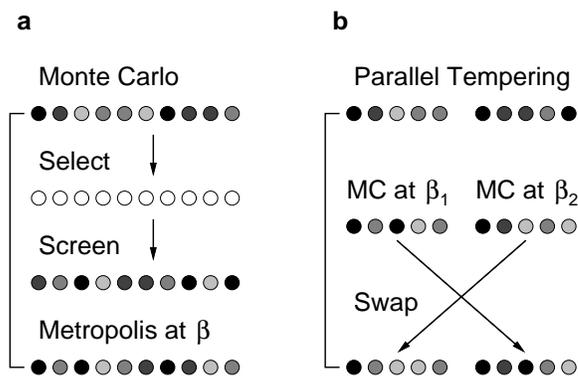,clip=,width=3.0in}
\caption{a) One Monte Carlo round with 10 samples.
 b) One Parallel Tempering round with 5 samples at $\beta_1$ and
5 samples at $\beta_2$.}
\label{fig1}
\end{figure}

Two ways of changing the variables are considered:
randomly changing the variables of a randomly
chosen sample a small
amount and exchanging a subset of
the variables between two randomly chosen samples.  These moves are
repeated until all the samples in a round have been modified.
The values of the figure of merit for the
proposed new samples are then measured.
Whether to accept the newly proposed
samples or to keep the current samples for the next round is decided
according to the
detailed balance acceptance criterion.  For the
random change of one sample, we find the Metropolis acceptance probability:
\begin{equation}
{\rm acc}(c \to p) = \min\left\{1,
\exp\left[\beta \left(E_{\rm proposed} - E_{\rm current }\right)\right]\right\} \ .
\label{2}
\end{equation}
  Proposed samples that
increase the figure of merit are always accepted; proposed samples that
decrease the figure of merit are accepted with the Metropolis
probability.  Allowing the figure of merit occasionally to decrease is
what allows samples to escape from local maxima.  
The random displacement of the $d$ mole fraction
variables, $x_i$, is done in the $(d-1)$-dimensional subspace orthogonal
to the $d$-dimensional vector
$(1,1,\ldots,1)$.  This procedure ensures that the constraint  $\sum_{i=1}^d x_i = 1$
is maintained.  This subspace is identified by the
Gram-Schmidt procedure.
Moves that violate the constraint $x_i \ge 0 $ are
rejected.  Moves that lead to invalid values of the non-composition variables
are also rejected.
For the swapping move applied to samples $i$ and $j$,
 we find the modified acceptance probability:
\begin{eqnarray}
{\rm acc}(c \to p) &=& \min\bigg\{1,
\exp\bigg[\beta \bigg(E_{\rm proposed}^i + E_{\rm proposed}^j
\nonumber \\
&&~~~~~~~~~~~~~~ - E_{\rm current}^i - E_{\rm current}^j \bigg)\bigg]\bigg\} \ .
\label{3}
\end{eqnarray}
Fig.~\ref{fig1}a
shows one round of a Monte Carlo procedure.  The parameter $\beta$
is not related to the thermodynamic temperature of the
experiment and
should be optimized for best efficiency.  The characteristic sizes of
the random changes in the composition and non-composition variables are also
parameters that should be optimized.

If the number of composition and non-composition variables is too great, or if
the figure of merit changes with the variables in a too-rough fashion,
normal Monte Carlo will not achieve effective sampling.  Parallel
tempering is a natural extension of Monte Carlo that is used to study
statistical \cite{Geyer}, spin glass \cite{Parisi}, and molecular \cite{Deem2}
systems with rugged energy landscapes.  Our most powerful
protocol incorporates the method of parallel tempering for changing
the system variables.  In parallel tempering, a fraction of
the samples are updated
 by Monte Carlo with parameter $\beta_1$, a fraction by
Monte Carlo with parameter $\beta_2$, and so on.  At the end of each
round, samples are randomly exchanged between the groups with
different $\beta$'s, as shown in Fig.~\ref{fig1}b.  
The acceptance probability for exchanging two samples is
\begin{equation}
{\rm acc}(c \to p) = \min\left\{1,
\exp\left[\Delta \beta \Delta E \right]\right\} \ ,
\label{4}
\end{equation}
where $\Delta \beta$ is the difference in the values of
 $\beta$ between the two groups,
and $\Delta E$ is the difference in the figures of merit between the
two samples.
It is important to
notice that this exchange step does not involve any extra screening compared
to Monte Carlo and is, therefore, ``free'' in terms of
experimental costs.  This step is, however, dramatically effective at
facilitating the protocol to escape from local maxima.  The number of
different systems and the temperatures of each system are parameters
that must be optimized.

To summarize, the first round of combinatorial chemistry consists of
the following steps:
constructing the  initial library of samples,
measuring the initial figures of merit,
changing the variables of each sample a small random amount or
swapping subsets of the variables between pairs of samples,
constructing the proposed new library of samples,
measuring the figures of merit of the proposed new samples, 
accepting or rejecting each of the proposed new samples, and
performing parallel tempering exchanges.
Following rounds of combinatorial chemistry repeat these steps,
starting with making changes to the current values
of the composition and non-composition
 variables.  These steps are repeated for as many rounds
as desired, or until maximal figures of merit are found.

We have chosen to sample the figure of merit by Monte Carlo, rather than
to optimize it globally by some other method, for several reasons.
First, Monte Carlo is
an effective stochastic optimization method.
Second, simple global optimization may be misleading since
concerns such as patentability, cost of materials, and ease of synthesis
are not usually included in the experimental figure of merit.
Moreover, the
screen that is most easily performed in the laboratory, the ``primary
screen,'' is usually only roughly correlated with the true figure of
merit.  Indeed, after finding materials that look
promising based upon the primary screen, experimental secondary and
tertiary screens are usually performed to identify that material which is
truly optimal.  Third, it might be advantageous to screen for several
figures of merit at once. For all of these reasons, sampling
by Monte Carlo to produce several candidate materials is
preferred over global optimization.


\section{The Random Phase Volume Model}

\begin{figure}[htbp]
\centering
\leavevmode
\psfig{file=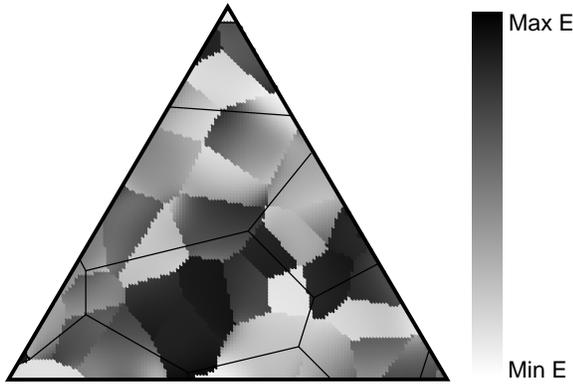,clip=,width=3.0in}
\caption{
The Random Phase Volume Model.  The model is shown for the
case of three composition variables and one non-composition variable.
The boundaries of the ${\bf x}$ phases are evident by the
sharp discontinuities in the figure of merit.  To generate this
figure, the ${\bf z}$ variable was held constant.  The boundaries of
the ${\bf z}$ phases are shown as thin dark lines.}
\label{fig2}
\end{figure}

The effectiveness of these protocols is demonstrated by combinatorial
chemistry experiments as simulated by
the Random Phase Volume Model.
The Random Phase Volume Model is not fundamental to the protocols;
it is introduced as a simple way to test, parameterize, and
validate the various searching methods.
The model relates the figure of merit to the composition and non-composition
variables in a statistical way.  The model is fast enough to allow for
validation of the proposed searching methods on an enormous number of
samples, yet possesses the correct statistics for the figure-of-merit
landscape.  The $d$-dimensional vector of composition mole
fractions is denoted by 
${\bf x}$. The composition mole fractions are non-negative and
sum to unity, and so the allowed compositions are constrained to lie within a 
simplex in $d$ dimensions.  For the familiar ternary system, this
simplex is an equilateral triangle.
  The composition variables are grouped into phases
centered around $N_x$ points ${\bf x}_\alpha$
randomly placed within the allowed composition range (the phases form
a Voronoi diagram \cite{Sedgewick}, see
Fig. \ref{fig2}).  The model is defined for any number of composition
variables, and the number of phase points is defined by requiring the
average spacing between phase points to be $\xi = 0.25$.  To avoid
edge effects, additional points are added in a belt of width $2 \xi$
around the simplex of allowed compositions.
The figure of merit should
change dramatically between composition phases.
Moreover, within each phase
$\alpha$, the figure of merit should also vary with ${\bf y} = {\bf x}
- {\bf x}_\alpha$ due to crystallinity effects such as crystallite
size, intergrowths, defects, and faulting \cite{8}. In addition, the
non-composition variables should
 also affect the measured figure of merit.  The
non-composition variables are denoted by the $b$-dimensional vector
${\bf z}$, with each component constrained to fall within the range
$[-1,1]$ without loss of generality.  
There can be any number of non-composition variables. 
The figure of merit depends on the composition and non-composition variables
in a correlated fashion, and so the non-composition variables 
also  fall within $N_z$ ``$z$-phases''
defined in the space of composition variables.  There are
a factor of 10 fewer non-composition phases than
composition phases.  The
functional form of the model when ${\bf x}$ is in 
composition phase $\alpha$ 
and non-composition-phase $\gamma$ is
\begin{eqnarray}
E({\bf x}, {\bf z}) &=& U_\alpha 
\nonumber \\
+ \sigma_x
\sum_{k=1}^q && \sum_{i_1 \ge \ldots \ge i_k = 1}^d
f_{i_1 \ldots i_k} \xi_x^{-k} A^{(\alpha k)}_{i_1 \ldots i_k} \, 
y_{i_1} y_{i_2} \ldots  y_{i_k}
\nonumber \\
 +  \frac{1}{2} {\Big (} W_\gamma  &&
\nonumber \\
+
\sigma_z
\sum_{k=1}^q && \sum_{i_1 \ge  \ldots \ge i_k = 1}^b
f_{i_1 \ldots i_k} \xi_z^{-k} B^{(\gamma k)}_{i_1 \ldots i_k} 
z_{i_1} z_{i_2} \ldots  z_{i_k} {\Big )},
\label{rpvm}
\end{eqnarray}
where $f_{i_1 \ldots i_k}$ is a constant symmetry factor, $\xi_x$ and
$\xi_z$ are constant scale factors, and
$U_\alpha$, $W_\gamma$, $A^{(\alpha k)}_{i_1 \ldots i_k}$, and
$B^{(\gamma k)}_{i_1 \ldots i_k}$ are random Gaussian variables with unit
variance. In more detail, the symmetry factor is given by
\begin{equation}
f_{i_1 \ldots i_k} = \frac{k!}{ \prod_{i=1}^l o_i!} \ ,
\label{6}
\end{equation}
 where $l$
is the number of distinct integer values  in the
set $\{i_1, \ldots, i_k\}$, and $o_i$
is the number of times that distinct value $i$ is repeated in the set.
Note that $ 1 \le l \le k$ and
$\sum_{i=1}^l o_i = k$.
The scale factors are chosen so that each term in the multinomial
contributes roughly the same amount: $\xi_x = \xi/2$ and $\xi_z = 
(\langle z^6 \rangle / \langle z^2 \rangle)^{1/4} =(3/7)^{1/4}$.
The $\sigma_x$ and $\sigma_z$ are chosen so that the
multinomial, crystallinity
 terms contribute 40\% as much as the constant, phase terms on
average. For both multinomials $q=6$.  
 As Fig.~\ref{fig2}
shows, the Random Phase Volume Model describes a rugged figure of
merit landscape, with subtle variations, local maxima, and
discontinuous boundaries.

\section{Results}
Six different search protocols are tested with increasing
numbers of composition and non-composition variables.
The total number of samples whose figure of merit will be measured is
fixed at $M=
100,000$, so that all protocols have the same experimental cost.
The single pass protocols Grid, Random, and LDS are considered.  
For the Grid method,
we define $M_x = M^{(d-1)/(d-1+b)}$ and $M_z =
M^{b/(d-1+b)}$.  The grid spacing of the composition variables
is
$\zeta_x = \left( V_d/M_x \right)^{1/(d-1)}$, where
\begin{equation}
V_d = \frac{\sqrt{d}}{(d-1)!}
\label{7}
\end{equation}
is the volume of the allowed composition simplex. Note that the
distance from the centroid of the simplex to the closest point on the
boundary of the simplex is 
\begin{equation}
R_d = \frac{1}{\left[d (d-1)\right]^{1/2}} \ .
\label{8}
\end{equation}
The spacing for each
component of the non-composition
 variables is $\zeta_z = 2/M_z^{1/b}$.
 For the LDS method, different quasi-random sequences are used for the
composition and non-composition variables.  
The feedback protocols Monte
Carlo, Monte Carlo with swap, and Parallel Tempering are 
considered.
The Monte Carlo parameters were optimized
on test cases.  It was optimal to perform
100 rounds of 1,000 samples with $\beta = 2$ for $d=3$ and
$\beta = 1$ for $d=4$ or 5, and
$\Delta x = 0.1 R_d$ and $\Delta z = 0.12$ for the maximum random
displacement in each component.
The swapping move consisted of an attempt to swap all of the non-composition
values between the two chosen samples, and it was optimal to use
$P_{\rm swap} \simeq 0.1$ for the
probability of a swap versus a regular random displacement.
 For Parallel Tempering it was optimal to perform 100 rounds with 1,000
samples, divided into three subsets:
50 samples at $\beta_1 = 50$, 500 samples at $\beta_2 = 10$, and 450 samples
at $\beta_3 = 1$.  The 50 samples at large $\beta$ essentially
perform a ``steepest-ascent'' optimization
 and have smaller $\Delta x = 0.01 R_d$
and $\Delta z = 0.012$.  

\begin{figure}[htbp]
\centering
\leavevmode
\psfig{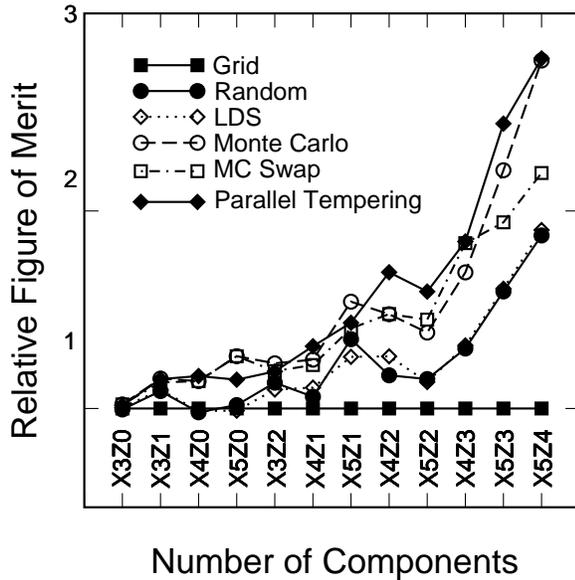}
\caption{The maximum figure of merit found with different protocols on
systems with different number of composition ({\bf x}) and non-composition
({\bf z}) variables. The results are scaled to the maximum found by
the Grid searching method. 
Each value is averaged over scaled results on 10
different instances of the Random Phase Volume Model with different
random phases.  
The Monte Carlo methods are especially
effective on the systems with larger number of variables, where the
maximal figures of merit are more difficult to locate.
}
\label{fig3}
\end{figure}

The figures of merit found by the protocols are 
shown in Fig.~\ref{fig3}. 
The Random and LDS protocols find
better solutions than does Grid in one round of experiment.
More importantly, the Monte Carlo methods have a tremendous advantage
over one pass methods, especially as the number of variables
increases, with Parallel Tempering the best method. 
The Monte Carlo methods, in essence, gather more information about
how best to search the variable space with each succeeding round.
This feedback mechanism proves to be effective even for the
relatively small total sample size of 100,000 considered here.
We expect that the advantage of the Monte Carlo methods  will become even
greater for larger sample sizes.   Note that in
cases such as catalytic activity, sensor response, or ligand
specificity \cite{Deem},
 the experimental figure of merit would likely be exponential
in the values shown in Fig.~\ref{fig3}, so that the success of the
Monte Carlo methods would be even more dramatic. A better calibration of
the parameters in Eq.~\ref{rpvm} may be possible as more data becomes
available in the literature.

\section{Conclusion}

To conclude, the experimental challenges in combinatorial chemistry
appear to lie mainly in the screening methods and in the
technology for the creation of the libraries.
The theoretical challenges, on the other hand,
appear to lie mainly in the library design and redesign strategies.  We
have addressed this second question 
via an analogy with Monte Carlo computer simulation, and
we have introduced the Random Phase Volume Model
to compare various strategies.
We find the multiple-round, Monte Carlo protocols to be especially
effective on the more difficult systems with larger numbers of 
composition and non-composition variables.  

An efficient implementation of the search strategy is feasible with
existing library creation technology.  Moreover
``closing the loop'' between library design and redesign
is achievable with the same database technology currently used
to track and record the data from  combinatorial chemistry experiments.
These multiple-round protocols, when combined with appropriate robotic
controls, should allow the practical
application of combinatorial chemistry to more complex and interesting
systems.

\section*{Acknowledgment}
This research was supported by the National Science Foundation
through grant number CTS--9702403.

\bibliography{combi}

\end{document}